\documentclass[preprint,12pt]{elsarticle}
\makeatletter
\def\ps@pprintTitle{%
 \let\@oddhead\@empty
 \let\@evenhead\@empty
 \def\@oddfoot{\textit{Submitted Preprint} \hfill}%
 \let\@evenfoot\@oddfoot}
\makeatother

\usepackage{amssymb}
\usepackage{amsmath}
\usepackage{amsthm}

\usepackage{amssymb}
\usepackage{amsmath}
\usepackage{color,soul}
\usepackage{bbm}
\usepackage{graphicx}     
\usepackage{nicematrix}
\usepackage{natbib}
\usepackage{adjustbox}
\usepackage{multirow}
\usepackage{array}
\usepackage{booktabs}
\usepackage{mathtools}
\usepackage{multicol}
\usepackage{algpseudocode}
\usepackage{algorithm}
\usepackage{xcolor}
\usepackage{caption}
\usepackage{pdflscape}

\newcommand{\argmax}{\operatornamewithlimits{argmax}}

\begin{document}

\begin{frontmatter}

\title{Enhanced Reinforcement Learning-based Process Synthesis via Quantum Computing}

\author[WVU]{Austin Braniff}\ead{austin.braniff@mail.wvu.edu}
\author[CU]{Fengqi You}\ead{fengqi.you@cornell.edu}
\author[WVU]{Yuhe Tian\corref{cor}}\ead{yuhe.tian@mail.wvu.edu}

\affiliation[WVU]{organization={Department of Chemical and Biomedical Engineering, West Virginia University},
            city={Morgantown},
            state={West Virginia},
            country={United States}}
\affiliation[CU]{organization={R.F. Smith School of Chemical and Biomolecular Engineering, Cornell University},
            city={Ithaca},
            state={New York},
            country={United States}}
\cortext[cor]{Corresponding author}

\begin{abstract}
In this work, we present quantum reinforcement learning (RL) as a solution strategy for process synthesis problems. Building on our prior work, we develop a generalized framework that formally poses process synthesis as a Markov decision process and introduces quantum-enhanced RL algorithms to solve it with improved scalability. Earlier implementations of quantum-based RL for process synthesis were limited by qubit requirements, which scaled poorly with problem complexity. This work overcomes this challenge by introducing state encoding algorithms to decouple qubit requirements from problem size. A classical RL-based solution strategy is used as a baseline to benchmark the quantum algorithms under identical training conditions. All algorithms are evaluated across a flowsheet synthesis problem of increasing unit counts to analyze their performance and scalability. Results show that all approaches are capable of identifying the optimal flowsheet designs in small design spaces. For moderate-scale unit counts, quantum approaches demonstrate competitive performance on a per-episode basis and improved efficiency on a per-parameter basis versus the classical RL benchmark. This work provides a foundation for future quantum computing applications within process systems engineering, establishes a controlled benchmark for comparing classical and quantum algorithms, and shows that the proposed quantum variants remain competitive for the process synthesis problem examined in this work.
\end{abstract}

\begin{keyword}
Quantum Computing \sep Reinforcement Learning \sep Process Synthesis \sep Process Design \sep Machine Learning


\end{keyword}

\end{frontmatter}



\section{Introduction}
Process synthesis remains a fundamental research topic in process systems engineering (PSE) by providing strategies to optimally design chemical and energy processes. This has become increasingly important as process industries strive to innovate designs that simultaneously meet economic, sustainable, and operability-based objectives \cite{pistikopoulosDataModelsAlgorithms2026,braniffRealtimeProcessSafety2025d}. Several solution strategies exist aiming to systematically solve these synthesis problems and provide optimal flowsheet configurations and operational variables \cite{mencarelliReviewSuperstructureOptimization2020}. Traditional formulations based on mixed-integer nonlinear programming (MINLP) \cite{boukouvalaGlobalOptimizationAdvances2016,linanTrendsPerspectivesDeterministic2025} or generalized disjunctive programming (GDP) \cite{grossmannSystematicModelingDiscretecontinuous2013} have enabled systematic optimization of predefined superstructures. Targeting approaches \cite{frumkinTargetBoundsReaction2018} and phenomena-based synthesis \cite{pistikopoulosSynthesisOperabilityStrategies2022,tianOverviewProcessSystems2018b} serve as alternative modeling frameworks that narrow design spaces and embed physical insight into the formulation \cite{pistikopoulosAdvancedModelingOptimization2024}. However, all of these approaches typically require explicit flowsheet information, rely on expert knowledge of the process, and/or become computationally demanding as combinatorial complexity grows. 

Reinforcement learning (RL) has recently emerged as a new technique for process flowsheet generation, in which process design is treated as a sequential decision-making problem rather than a large-scale static program \cite{gaoDeepReinforcementLearning2024e}. In this setting, an agent incrementally modifies a flowsheet representation and evaluates performance through simulation-based rewards \cite{gottlAutomatedSynthesisSteadystate2022}. By avoiding rigid superstructure assumptions, RL offers a flexible mechanism for exploring design alternatives. Nevertheless, these deep RL methods face practical limitations when applied to large process design spaces, including slow convergence, sensitivity to hyperparameters, and increasing difficulty in efficiently representing high-dimensional state-action value functions.

Quantum computing has been proposed as a new computing strategy with the potential to accelerate classical algorithms \cite{bernalPerspectivesQuantumComputing2022b,ajagekarNewFrontiersQuantum2022c}. In PSE, quantum computing has already improved many subfields, such as: optimization \cite{bernalIntegerProgrammingTechniques2020,ajagekarQuantumComputingBased2020}, control \cite{niemanControlImplementedQuantum2022}, and simulation \cite{heinenQuantumComputingComplex2024}. Towards the direction of reinforcement learning, the quantum computing domain contains machine learning-analogous models called parameterized quantum circuits (PQCs) for function representations and approximations \cite{benedettiParameterizedQuantumCircuits2019a}. In particular, these quantum circuits have been investigated as parameterized models that may serve as value-function approximators within reinforcement learning algorithms \cite{meyerSurveyQuantumReinforcement2024}, potentially leading to agents that can learn faster and explore more efficiently. While current hardware constraints limit problem size, hybrid quantum–classical algorithms provide a platform for studying whether quantum representations can reliably improve learning behavior in structured combinatorial problems such as process synthesis.

\begin{figure}[!h]
    \centering
    \includegraphics[width=\linewidth]{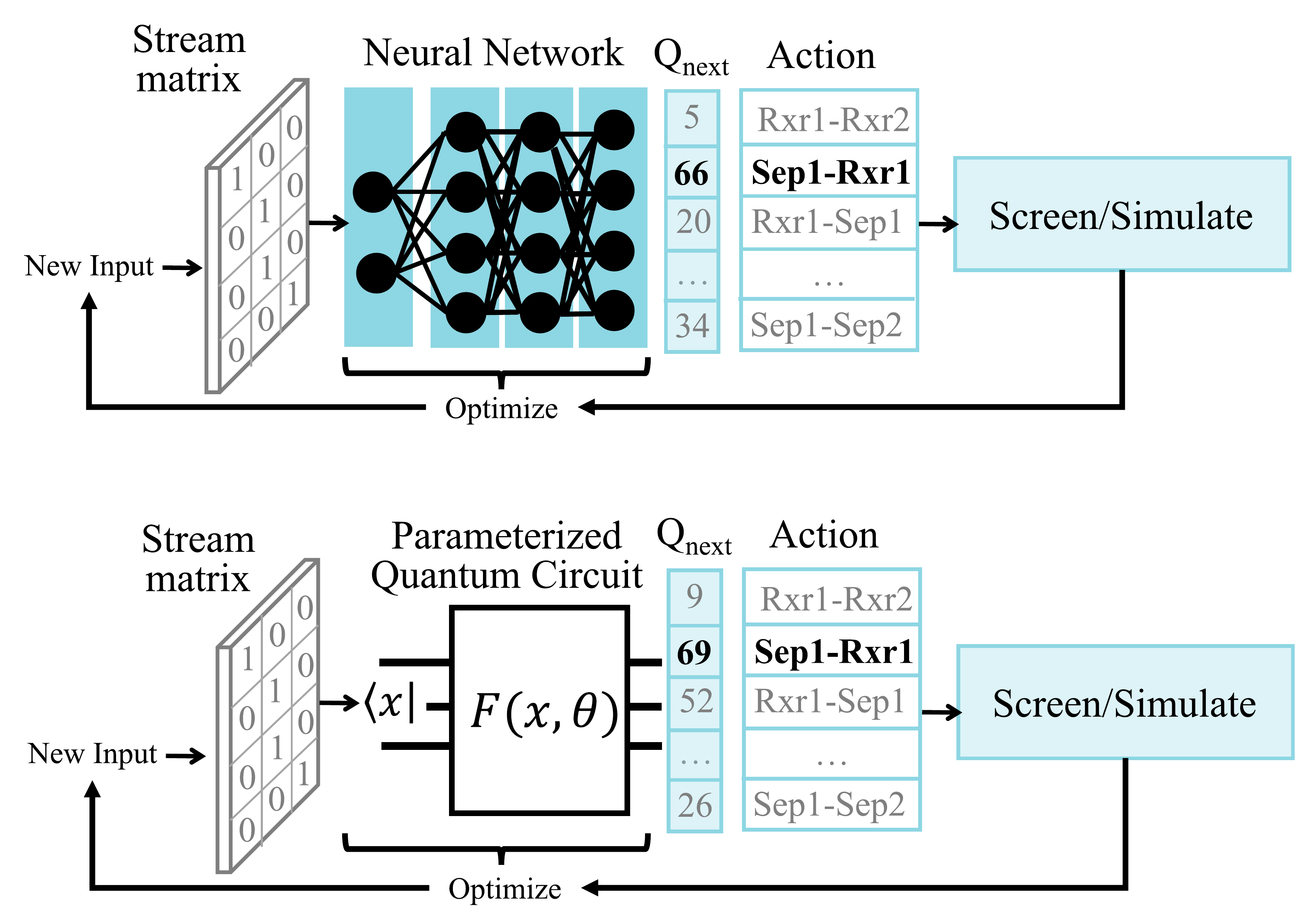}
    \caption{An overview of the classical and quantum RL approaches for process design.}
    \label{fig:framework}
\end{figure}

In our earlier work, we investigated preliminary integrations of quantum models within reinforcement learning for simplified flowsheet problems \cite{braniffEnhancedReinforcementLearningdriven2025d} which highlighted both the potential and the limitations of direct quantum implementations, particularly with respect to qubit scaling and encoding strategies. The present work advances that foundation by developing a generalized framework with improved scalability for quantum-enhanced reinforcement learning in process design and benchmarking it against classical RL approaches, as depicted in Fig. \ref{fig:framework}. This paper is an extended version based on our conference proceeding submitted to The 36th European Symposium on Computer Aided Process Engineering (ESCAPE-36) \citep{braniffProcess2026}. We formalize the process synthesis problem as a Markov Decision Process (MDP), establish a consistent classical deep Q-network benchmark, and systematically evaluate multiple quantum encoding strategies under a unified formulation. The objective of this study is therefore not only to demonstrate feasibility but also to examine how quantum-enhanced reinforcement learning behaves as process complexity increases. Specifically, we aim to (i) define a rigorous MDP formulation for flowsheet synthesis, (ii) compare classical and quantum value-function approximators under consistent training conditions, and (iii) analyze scaling trends and limitations of quantum computing for process synthesis problems. 

The remaining sections of this paper are organized as follows: Sections 2 and 3 provide the background information necessary for the reinforcement learning and quantum computing component of this work. Section 4 establishes a formal definition for process synthesis as an MDP and describes our proposed quantum-based algorithm along with three variants. Section 5 demonstrates and discusses the viability and scalability of these algorithms on a representative case study with various degrees of complexity. Section 6 gives the conclusions and presents ongoing research efforts.  

\newpage
\section{Reinforcement Learning}
\begin{figure}[!h]
    \centering
    \includegraphics[width=0.8\linewidth]{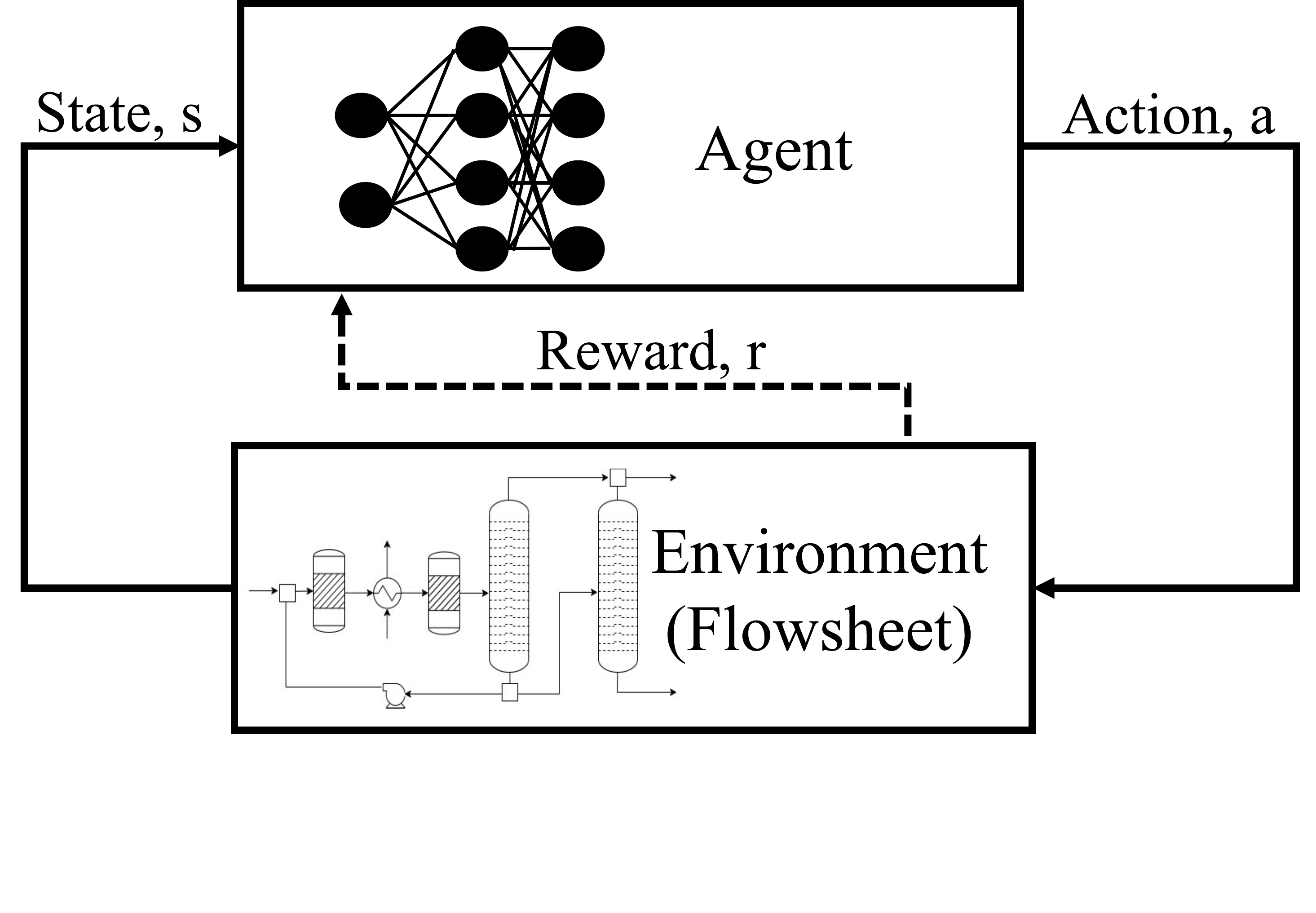}
    \caption{Value-based deep reinforcement learning nomenclature for the present work.}
    \label{fig:RL}
\end{figure}

Reinforcement learning (RL) is typically introduced as an algorithm to solve stochastic dynamic optimization problems, more commonly referred to as Markov decision processes (MDPs). In the present work, we will apply reinforcement learning towards flowsheet synthesis problems after framing them as MDPs. There are three essential components in RL that describe the dynamics between an agent and the environment: (i) the state of the environment, $s$, which characterizes all relevant information about the environment, (ii) the action being taken onto the environment, $a$, which is determined by the RL agent, and (iii) the immediate reward from the environment, $r$, which is a feedback signal designed to encourage desirable behavior from the environment. The interplay of these elements is depicted within an RL framework in Fig. \ref{fig:RL}. Based on these variables, two kinds of functions can be defined, which are necessary for modern RL algorithms. These are the state-value function, $V$, and the state-action value function, $Q$, and they are given in Eqs. \ref{eq:Value_fun} and \ref{eq:Q_fun} below where $a_t^\star$ represents an optimal action and $\gamma$ represents a discount factor. 

\begin{equation}
\label{eq:Value_fun}
    V(s_0) =  \mathbb{E}[\sum_{t=0}^\infty \gamma^t r(s_t, a_t^\star)]
\end{equation}
\begin{equation}
\label{eq:Q_fun}
    Q(s_0,a_0) = r(s_0,a_0) +  \mathbb{E}[\sum_{t=1}^\infty \gamma^t r(s_t, a_t^\star)]
\end{equation}

The value function represents the expected discounted sum of the reward for all time, assuming that optimal actions are taken for all time steps, including the current one. In this way, it serves as a representation of how desirable it is for the environment to exist in a given state. The state-action value function is slightly more complex but follows a similar principle. It represents the expected discounted sum of the reward for all time after taking a specific action and the following optimal actions afterward. From this, the state-action value function provides the desirability of choosing a specific action in addition to existing in a particular state. These functions are necessary in modern RL as they are usually approximated using neural networks (NNs) and then used to guide an agent's policy updates. For a more thorough introduction into RL, the reader is referred to Sutton and Barto \cite{sutton2018reinforcement}. 

\subsection{RL for process synthesis}
While reinforcement learning has been utilized in process control \cite{shinReinforcementLearningOverview2019a,braniffReinforcementLearningbasedControl2026} and operational decision-making \cite{rangel-martinezInterpretableOnlineScheduling2026,hubbsDeepReinforcementLearning2020}, its application to structural process design remains comparatively recent. Process synthesis problems are inherently combinatorial. The selection, arrangement, and interconnection of unit operations, along with the determination of optimal operating variables, define a large design space that grows rapidly with the number of candidate processing units. Reinforcement learning solves these problems in a sequential and iterative fashion. Several representations have been proposed to encode flowsheet structures for learning-based approaches, including graph-based \cite{stopsFlowsheetGenerationHierarchical2023a,GAO20232005} and matrix-based encodings \cite{khanSearchingOptimalProcess2020,kimProcessDesignOptimization2023}. In these formulations, binary or integer-valued variables describe the connectivity between candidate unit operations, providing a compact state representation that can be processed by neural networks. The action space typically corresponds to permissible structural modifications subject to feasibility constraints, such as avoiding invalid unit connections or ensuring physically meaningful sequencing of operations \cite{wangReinforcementLearningAutomated2022}. Through this formulation, an agent incrementally modifies a flowsheet by selecting discrete actions such as adding, removing, or reconfiguring unit operations. Rather than enumerating all possible structures in advance, the agent explores the design space through interaction with a simulation-based environment and learns policies that favor high-performing configurations. In addition to various flowsheet representations, different RL-based algorithms have been proposed to solve various levels of process synthesis problems. These include solving for discrete decision variables only \cite{gottlAutomatedFlowsheetSynthesis2021a,GOTTL20221555}, selecting continuous and discrete variables concurrently \cite{reynoso-donzelliReinforcementLearningApproach2025,seidenbergBoostingAutonomousProcess2023}, simultaneous design and control \cite{sachioIntegratingProcessDesign2022,reynoso-donzelliIntegratedReinforcementLearning2025c}, or leveraging transfer learning to avoid expensive simulations \cite{gaoAcceleratingProcessSynthesis2025c,GAO20232005}.

\subsection{Deep Q-Learning}
In the present study, the deep Q-network (DQN) algorithm \cite{mnihPlayingAtariDeep2013} serves as the classical benchmark for comparison against quantum-enhanced value-function approximators. The DQN algorithm is an NN-based extension of Q-learning and represents one of the foundational algorithms in model-free RL. Value-based RL is used over policy-gradient or actor-critic methods because the discrete action space in process synthesis naturally allows for direct action value assignment over the finite decision space. In Q-learning, the objective is to approximate the optimal action-value function $Q^*(s,a)$, which represents the maximum expected cumulative discounted reward obtained by taking a particular action $a$ in a particular state $s$ and following the optimal policy for all time after. As the dimensionality of the state-action space increases, directly storing or computing Q-values becomes computationally infeasible. DQN addresses this limitation by parameterizing the Q-function using a neural network, yielding an approximation of the form $Q_{\theta}$ where $\theta$ represents the trainable NN parameters. The theoretical foundation of Q-learning is the Bellman optimality equation, as seen in Eq. \ref{eq:bellman} below. 

\begin{equation}
\label{eq:bellman}
    Q(s,a) = r+ \gamma \max_{a_{t+1}}Q(s_{t+1},a_{t+1})
\end{equation}
where $r$ is the reward for a given state-action pair, $\gamma \in [0,1]$ is a discount factor, and $s_{t+1}$ and $a_{t+1}$ denote the next state and candidate future actions, respectively. This recursive equation relates the relative value of a current state-action pair to the expected value of future decisions and forms the objective to iteratively improve through experience. In the DQN algorithm, both sides of Eq. \ref{eq:bellman} are approximated by NNs. These two networks are termed the evaluation network $Q_{\theta}(s,a)$ and the target network, $Q_{\theta^-}(s,a)$. The evaluation network is typically updated at every training step while the target network is updated periodically to become a copy of the evaluation network. The target network with its periodic updating provides a stable reference on the right side of the Bellman optimality equation. At every step, the parameters of the evaluation network are selected by performing one step of gradient descent on the mean squared error between the predicted Q-value and its target as shown in Eq. \ref{eq:MSE}. 

\begin{equation}
\label{eq:MSE}
    \min_{\theta} \mathcal{L}(\theta) = \mathbb{E}[(r+\gamma \max_{a_{t+1}}Q_{\theta^-}(s_{t+1},a_{t+1}) - Q_{\theta}(s,a))^2]
\end{equation}

The DQN algorithm has two more critical features which enable a stable yet effective learning algorithm. The first is experience replay where instead of updating the network using consecutive state transitions, interactions are stored in a replay buffer and sampled randomly in mini-batches. This technique avoids the temporal correlations that may be present in consecutive samples and eliminates biasing the learning to more recent agent experiences. The second is an $\epsilon$-greedy exploration policy for the action selection. Under this policy, the action is chosen with probability $\epsilon$ to be a random action within the action space or to be the greedy action following $a = \argmax_aQ(s,a)$ otherwise. The value of $\epsilon$ is typically small and decays to zero over time. This allows greater exploration in the early episodes of training and greater exploitation in the later episodes.

\section{Quantum Computing}
\begin{figure}[!h]
    \centering
    \includegraphics[width=0.5\linewidth]{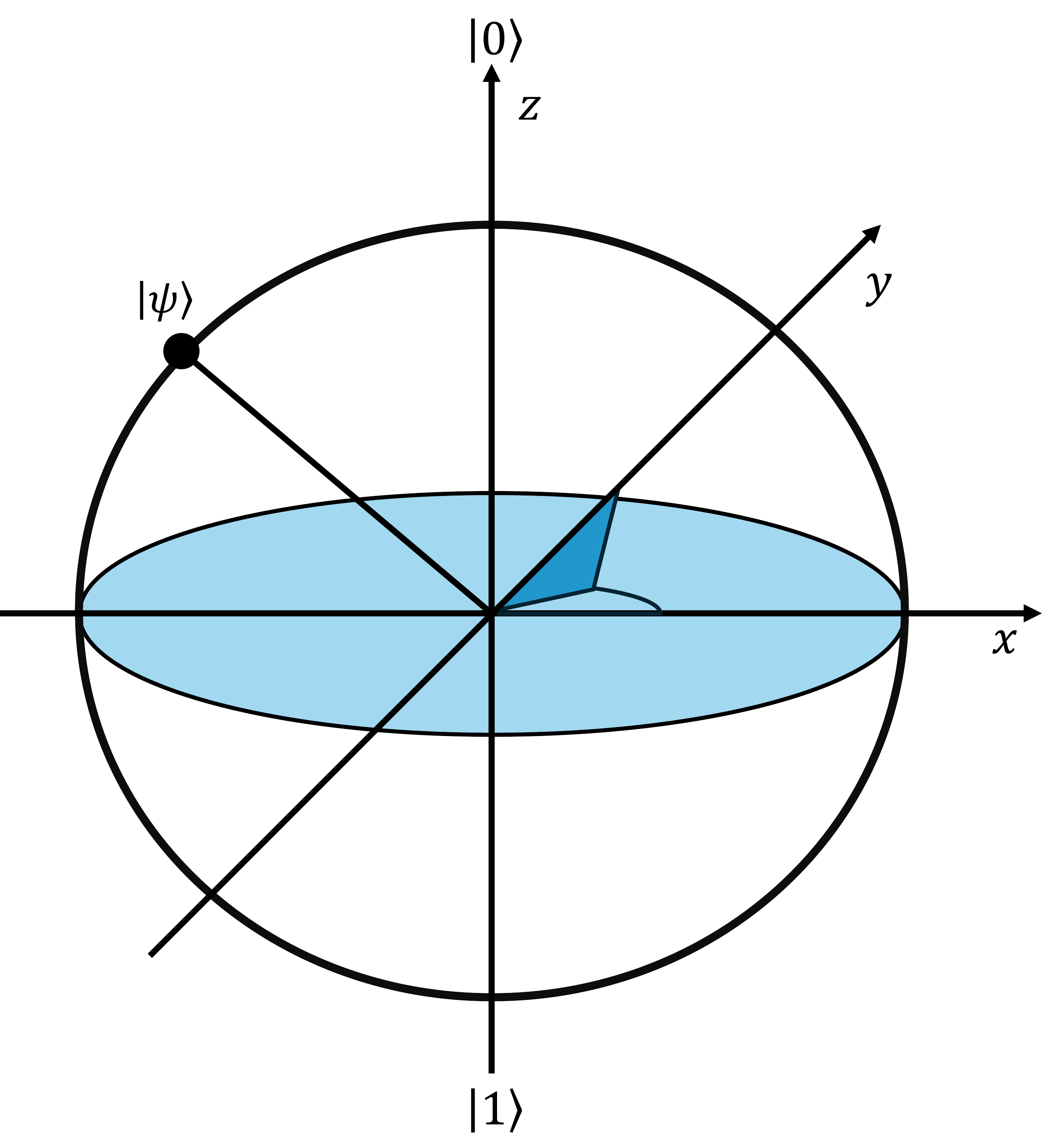}
    \caption{Bloch sphere diagram for quantum state representation.}
    \label{fig:bloch}
\end{figure}

Quantum computing is based on the manipulation of quantum bits, or qubits, which differ fundamentally from classical binary bits. A classical bit exists in one of two discrete states (0 or 1). A qubit is a bit more complex, but ultimately it can take one of two computational basis states when measured. In ket notation these states are represented as $|0\rangle$ and $|1\rangle$ corresponding to the value of classical binary bits. What essentially separates qubits from bits is the phenomenon that qubits can take a superposition of the computational basis states. This superposition is represented by a two-dimensional state vector, $|\psi\rangle$, belonging to a Hilbert space and is expressed as written in Eq. \ref{eq:hilbert}.
\begin{align}
\label{eq:hilbert}
    &|\psi\rangle = \alpha|0\rangle+\beta|1\rangle\\
    \text{where} &\quad |\alpha|^2+|\beta|^2=1 \nonumber
\end{align}
$\alpha$ and $\beta$ are complex probability amplitudes representing the likelihood that the qubit will be measured in state $|0\rangle$ and state $|1\rangle$, respectively. The direct probability is represented by the squared magnitudes of each $\alpha$ and $\beta$, for example, the probability that a qubit will be measured in state $|0\rangle$ is defined by $|\alpha|^2$. The state of a qubit is geometrically represented by a vector from the origin to a point exisiting on the surface of the Bloch sphere shown in Fig. \ref{fig:bloch}. This sphere is defined by three orthogonal axes, x, y, and z, corresponding to Pauli operators.

For an $n$-qubit system, the basis state grows exponentially to $2^n$ representing all possible combinations of measuring each qubit in a $|0\rangle$ or $|1\rangle$ state. In this way, superposition is expressed over exponentially many basis states. Operations acting on multiple qubits can produce entangled states, in which the state of one qubit cannot be explicitly described independently of the others. This conditioning is described as entanglement, and it is a key feature exploited in quantum algorithms and quantum models. Quantum computation proceeds through the application of unitary operations, implemented as quantum gates. Single qubit gates usually correspond to rotations on the Bloch sphere, which are parameterized by rotation angles. These gates are represented by $R_X$, $R_Y$, and $R_Z$ and represent rotations around the $x$, $y$, and $z$ axis of the Bloch sphere (Fig. \ref{fig:bloch}), respectively. Multi-qubit gates, such as controlled operations (e.g., controlled-Z ($CZ$) gates), introduce correlations between qubits and can impose quantum entanglement. Multi-qubit systems that use these gates are more commonly referred to as quantum circuits. These quantum circuits are specifically designed for the task they are being used for, where particular circuit architectures are best suited to solve particular problems. In these circuits, the qubits are initialized to a certain state representation and then acted on by quantum gates, and then the qubit states or specific physical properties (observables) are measured, providing a single computational basis state as a result for the circuit. For a deeper introduction into quantum computing, the reader is referred to Harwood et al. \cite{harwoodFormulatingSolvingRouting2021} and Raseena \cite{raseenaQuantumComputingFoundations2025}. 
\subsection{Quantum Machine Learning and Reinforcement Learning}
Quantum machine learning (QML) investigates the use of quantum computing architectures to implement or enhance machine learning models. Unlike classical neural networks, quantum circuits do not directly output deterministic values. Instead, information is extracted through measurement. Observables are measured repeatedly to estimate expectation values, and these estimates can require repeated circuit executions for reliable results. The need for sampling introduces statistical uncertainty and contributes significantly to both computational and practical costs, particularly for deep circuits or those with a large number of qubits. Current quantum hardware operates in what is described as the noisy intermediate-scale quantum (NISQ) regime \cite{preskillQuantumComputingNISQ2018}, characterized by limited qubit counts, quantum states changing with time, and imperfect gate implementations \cite{wangComprehensiveReviewQuantum2024}. As a result, circuit depth and qubit usage must be carefully managed in practical implementations. For these reasons, quantum machine learning at the current scale usually adopts hybrid quantum–classical schemes in which quantum circuits provide parameterized function approximations while parameter optimization is performed classically. Since the use of real-world quantum machines can become costly, most examples of QML algorithms use a quantum computing simulator to represent the quantum circuit \cite{biamonteQuantumMachineLearning2017,chenDesignAnalysisQuantum2024}. 

\begin{figure}[!h]
    \centering
    \includegraphics[width=0.9\linewidth]{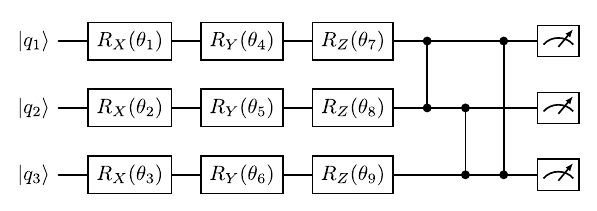}
    \caption{Circuit diagram of a PQC.}
    \label{fig:PQC}
\end{figure}

The central modeling component in QML is the parameterized quantum circuit (PQC), while the full algorithm and optimization process of training the PQC is referred to as a variational quantum circuit (VQC) \cite{benedettiParameterizedQuantumCircuits2019a,schuldMachineLearningQuantum2021}. A PQC consists of layers of quantum gates with adjustable parameters (e.g., rotation gates with tunable rotation angles) that are optimized with respect to a defined loss function. From a modeling perspective, PQCs serve as trainable function approximators. The circuit parameters are updated using classical gradient-based or gradient-free optimizers, similar to weight updates in neural networks. Although the mathematical structure differs from classical networks, both approaches aim to approximate nonlinear mappings between inputs and outputs through tunable parameters. A diagram of a 3-qubit PQC is given in Fig. \ref{fig:PQC} where $q_i$ represents the $i^{th}$ qubit, $R_j$ is a rotation gate for the $j$-axis, $\theta$ is a set of tunable parameters, a connection between qubits represents a CZ gate, and the meter represents measuring the end-state of a qubit.

\begin{figure}
    \centering
    \includegraphics[width=0.8\linewidth]{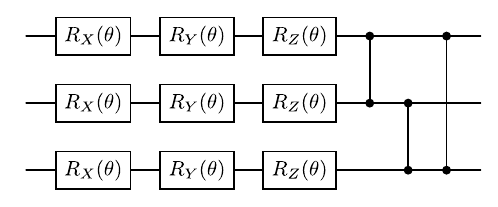}
    \caption{Circuit diagram for a PQC layer.}
    \label{fig:PQC_layer}
\end{figure}

Classical data may be encoded into the circuit through gate rotations or quantum state initializations, and model outputs are obtained by measuring observables at the end of the circuit. A typical PQC layer includes single qubit rotation gates followed by entangling gates such as CZ gates. The rotation gates apply parameterized transformations about the corresponding axes of the Bloch sphere, while entangling gates introduce correlations between qubits. In this work, a sequence of parameterized single-qubit rotations followed by entangling operations is referred to as one PQC layer, as shown in Fig. \ref{fig:PQC_layer}.

\begin{figure}[!h]
    \centering
    \includegraphics[width=0.9\linewidth]{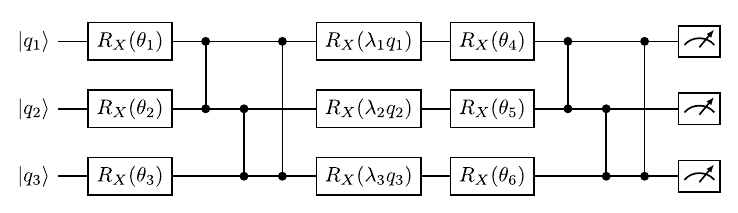}
    \caption{Circuit diagram for a re-uploading PQC.}
    \label{fig:Reup_PQC}
\end{figure}

An important modeling concept in QML is data re-uploading. Instead of encoding input features only once at the beginning of a circuit, data can be reintroduced at multiple layers through additional rotation gates where the angles depend on the input data. When using this within a PQC the circuit is then referred to as a re-uploading PQC. An example of a re-uploading PQC is provided in Fig. \ref{fig:Reup_PQC} where $\lambda$ is a set of tunable scaling parameters acting on the original inputs embedded into the qubit states.

Data re-uploading increases the effective expressivity of the model and may allow more information to be embedded without increasing the number of qubits. This concept is particularly relevant in resource-constrained settings where qubit counts or circuit depth must remain limited. In this study, we will rely on different data encoding strategies to mitigate computational costs. 

For reinforcement learning applications, PQCs can replace neural networks as value-function or policy approximators \cite{chenIntroductionQuantumReinforcement2024,chenVariationalQuantumCircuits2020}. In value-based methods such as DQN, the PQC serves as a parameterized mapping from state representations to action-value estimates. Training proceeds through classical optimization of circuit parameters using reward-driven loss functions, while circuit evaluations are performed via quantum simulation or hardware execution. Aside from encoding and measurement considerations, the surrounding reinforcement learning framework remains unchanged. To this end, augmenting existing deep RL strategies is as simple as swapping neural networks with their quantum counterpart, the PQC. Quantum RL is relatively new being developed in the last few years, however it has already shown promise in various applications such as solving maze problems \cite{chenDeepQLearningHybrid2023}, the classic cart pole problem \cite{skolikQuantumAgentsGym2022b}, and other computing benchmarks such as MountainCar and Acrobot \cite{jerbiParametrizedQuantumPolicies2021b}. These developments motivate the investigation of quantum value-function approximators for combinatorial process synthesis problems, as examined in this work.

\section{Methodology}
\subsection{Process design as an MDP}
As briefly mentioned in Section 2, RL algorithms are conventionally used for solving dynamic or sequential decision-making problems. In this way, it may not be apparent how one can formulate a process design problem as a sequential decision-making task. The process synthesis problem is usually formulated as an MINLP or a GDP, and all the optimal design and operating parameters are determined by solving a single optimization problem. In contrast, when applying RL towards process synthesis, the set of optimal decision variables is determined in a step-by-step fashion. Instead of knowing the full set of possible combinations at once, the RL agent only observes a flowsheet in its current state and then adds or removes a unit operation giving a new flowsheet. This iterative process aims to build the optimal flowsheet representation over many iterations of modifying and simulating the process. 

\begin{figure}[!h]
    \centering
    \includegraphics[width=\linewidth]{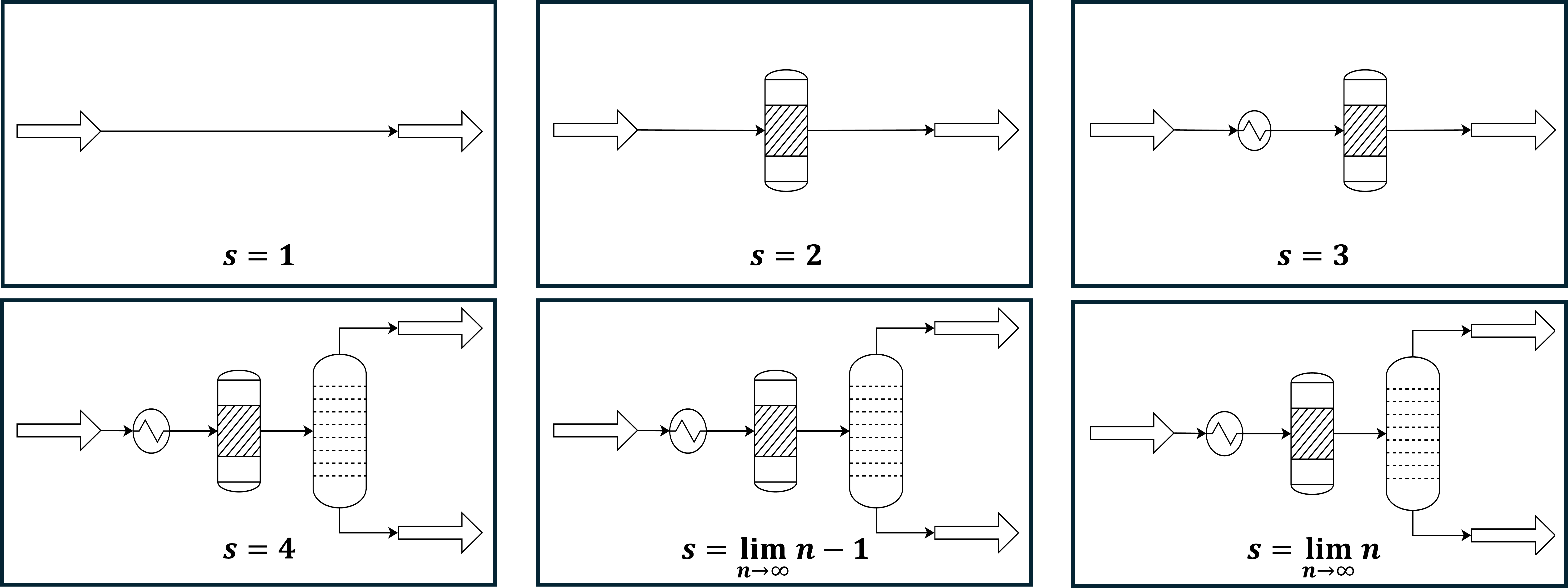}
    \caption{Flowsheet configurations following an optimal policy for an infinite horizon.}
    \label{fig:inf_horizon}
\end{figure}

The underlying process synthesis problem can be posed as an infinite horizon discounted MDP as shown below in Eq. \ref{eq:MDP}.
\begin{align}
\label{eq:MDP}
\pi^* & = \max_\pi \mathbb{E}[\sum_{t=0}^{\infty}\gamma^tr(s_t,a_t)]\\
\text{s.t.}\quad & s_{t+1} \sim P(s_{t+1}|s_t,a_t) \nonumber \\
&a_t = \pi(s_t) \nonumber
\end{align}
\noindent where $\pi$ is the deterministic policy that computes actions $a_t$ given the current state of the environment $s_t$. In the context of process synthesis, $s_t$ is the current state of the flowsheet as defined by which unit operations are present and their specific structural arrangement. The state space is defined by all possible flowsheet representations, this inherently includes the discrete combinatorial design space present in the formulation of MINLPs and GDPs. $a_t$ is the action taken by following a policy, $\pi$, and it consists of manipulating the flowsheet. The action space is defined by all possible single manipulations that can be taken onto the flowsheet (e.g., adding/removing a specific unit) and includes the ability to make no manipulation at all. $r$ is a predefined reward function based on design objectives of the process (e.g., cost, sustainability, productivity, etc.). The discount factor, $\gamma$, is assigned a value in the interval $(0,1)$ to ensure bounded convergence of the infinite sum. $P$ is the state transition function that is conditioned only on the current state and current action. The objective of this problem is to find an optimal policy, $\pi^*$, that maximizes the reward function (i.e., the process design objective) over the infinite horizon. It is important to note that making no manipulation to the flowsheet at a given step is a valid potential action from the policy. Therefore, once the optimal design is created, the theoretically optimal policy would not make any manipulation for all remaining steps within the infinite horizon. This is conceptually represented in Fig. \ref{fig:inf_horizon}, where the number of steps approaches infinity and the optimal configuration is found at step $s=4$.

\subsection{Classical RL baseline}
This section presents the classical RL-based process synthesis algorithm used as a benchmark for comparison with the proposed quantum approaches. The classical framework adopts the established RL-based process synthesis formulations \cite{wangReinforcementLearningAutomated2022} and our prior work \cite{tianReinforcementLearningDrivenProcess2024d}, serving as a reference point to assess the relative strengths and limitations of quantum-enhanced strategies. The benchmark employs the DQN algorithm \cite{mnihPlayingAtariDeep2013} to solve the process synthesis MDP problem. The full implementation is summarized in Algorithm 1.

\begin{algorithm}[htbp!]
\caption{DQN for Process Synthesis}
    \begin{algorithmic}[1]
        \State Input $\mathcal{S}$ with size $n$, a user-defined set of candidate unit operations
        \State Initialize replay memory $\mathcal{B}$ with capacity $N$
        \State Initialize Q-network, $Q_{\theta}$, with random weights $\theta$
        \State Initialize target Q-network, $Q_{\theta^-}$ with weights $\theta^{-} = \theta$
        \For{episode $= 1$ \textbf{to} $M$}
            \State Initialize a blank flowsheet $s_0$
            \For{$t = 1$ \textbf{to} $T$}
                \State With probability $\epsilon$, select a random action $a_t$
                \State Otherwise select $a_t = \argmax_a Q_{\theta}(s_t, a)$
                \State Modify flowsheet according to action $a_t$ to generate flowsheet $s_{t+1}$
                \If{flowsheet $s_{t+1}$ passes screening}
                \State Simulate flowsheet to observe reward $r_t$
                \State Store transition $(s_t, a_t, r_t, s_{t+1})$ in $\mathcal{B}$
                \State Sample random batch of transitions $(s_j, a_j, r_j, s_{j+1})$ from $\mathcal{B}$
                \State Compute target action:
                \[
                a_{j+1} = \argmax_a Q_{\theta^-}(s_{j+1},a)
                \]
                \State Compute target:
                \[
                    y_j = r_j + \gamma \max_{a'} Q_{\theta^-}(s_{j+1}, a')
                \]
                \State Perform a gradient descent step on $(y_j - Q_{\theta}(s_j, a_j))^2$ with respect to $\theta$
                \Else{ Continue}
                \EndIf 
                \If{every $C$ steps}
                    \State Update target network: $\theta^{-} = \theta$
                \EndIf
                \State $s_t \gets s_{t+1}$
            \EndFor
        \EndFor
    \end{algorithmic}
\end{algorithm}

In this algorithm, the process flowsheet is represented by a matrix of binary variables with dimensions equal to the number of available candidate units. That is, $s \in \{0,1\}^{n\times n}$ where $n$ denotes the number of available unit operations. Each non-zero element $s_{ij}$ indicates a connection from the outlet of unit $j$ to the inlet of unit $i$. The feeds and products of the process are also modeled as unit operations in this sense. This provides a compact encoding strategy of the flowsheet configuration at any given decision step. The action space encompasses a set of discrete manipulations that can be made to the flowsheet, such as the addition or removal of unit operations. Each action corresponds to a modification of one element of the stream connectivity matrix. The number of potential actions scales proportionally to the number of potential stream connections which grows combinatorially with the problem size.

\begin{figure}[h]
    \centering
    \includegraphics[width=0.7\linewidth]{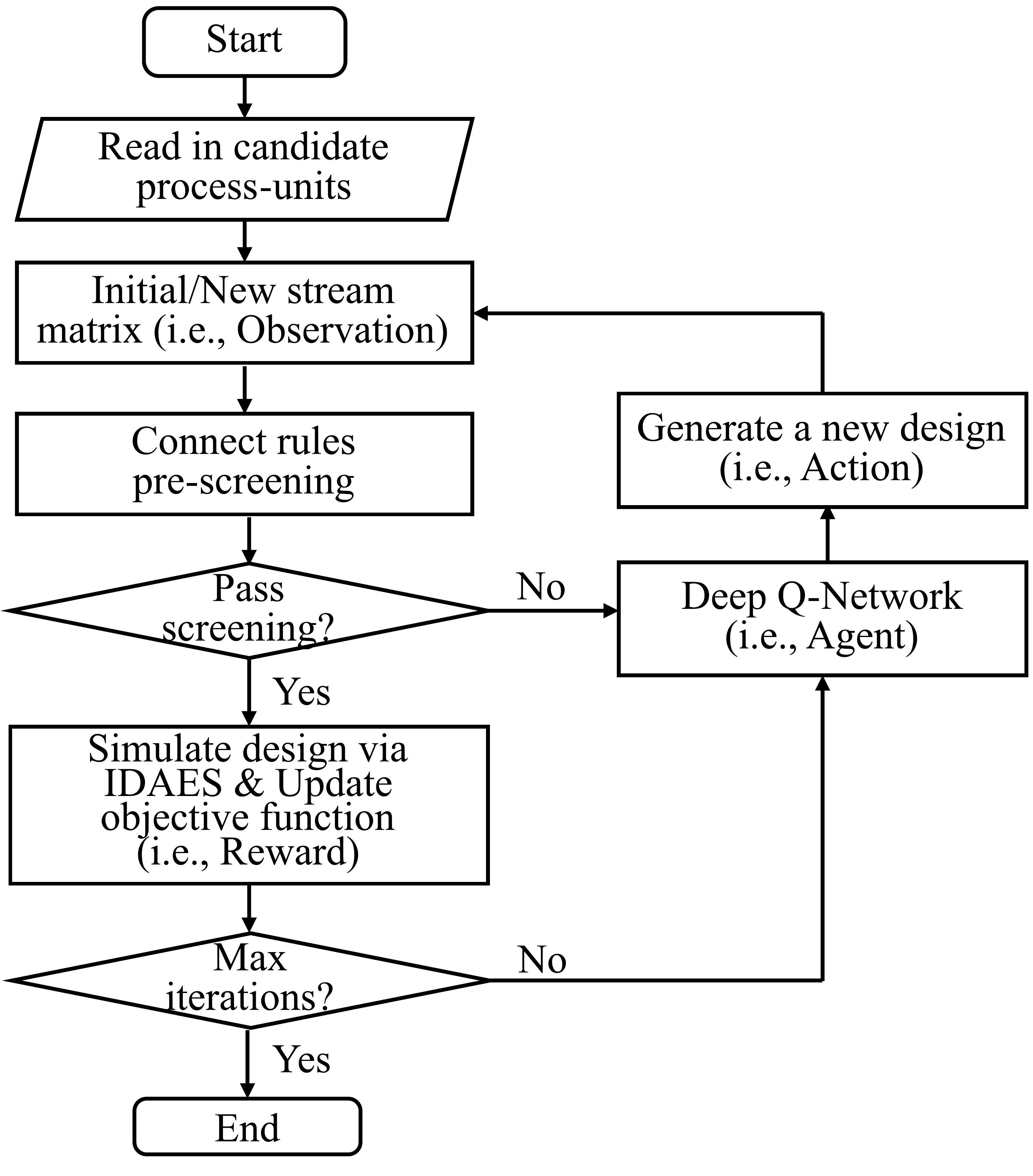}
    \caption{Workflow for RL-driven process synthesis (adapted from Wang et al. \cite{wangReinforcementLearningAutomated2022}).}
    \label{fig:procedure}
\end{figure}

As shown in Fig. \ref{fig:procedure}, at each step, a feasibility screening procedure is applied prior to simulation. This screening enforces structural and physical constraints, such as ensuring the flowsheet has an inlet and an outlet, invalid unit connections are avoided (e.g., a compressor directly feeding to an expander), and logical conditions are met (e.g., at least one reactor is used if there is a species change from the inlet to the outlet). If a proposed modification violates feasibility constraints, a penalty is assigned to the reward and the simulation step is skipped. Otherwise, the resulting flowsheet is passed to the IDAES-PSE \cite{leeIDAESProcessModeling2021} modeling environment, where continuous design and operating variables are optimized using IPOPT \cite{bieglerLargescaleNonlinearProgramming2009}.

Under this structure, the reinforcement learning component addresses the discrete structural decisions, while deterministic nonlinear optimization resolves the continuous operating variables. This decomposition allows each computational layer to handle the variable types most suited to its strengths. All optimization is performed with respect to a user-defined reward function, which may incorporate economic, environmental, or process performance objectives. The reward formulation is fully general and may include nonlinear or multi-objective components, provided they can be evaluated for a given flowsheet configuration. More details on this classical RL-based process synthesis algorithm can be found in Wang et al. \cite{wangReinforcementLearningAutomated2022}.

\subsection{Quantum RL-based process design}
Quantum RL-based process design is the use of quantum RL algorithms to solve process design problems. In this work, we augment the process synthesis DQN algorithm into a quantum-DQN algorithm for process synthesis. A comparison between the two algorithms demonstrates how simple it is to retrofit classical RL algorithms into quantum-enhanced versions. Nevertheless, the scalability and resource requirements of PQCs depend on qubit number, circuit depth, and measurement sampling. These constraints motivate careful architectural design when comparing quantum and classical function approximators, particularly in structured combinatorial problems such as process synthesis.

\subsubsection{Encoding and decoding}
As mentioned in Section 3, the encoding and decoding of classical data into a quantum format is a necessary step when employing quantum machine learning on classical data. In this work, the encoding and decoding strategies are intentionally made to be simple, as the process synthesis problem involves data that can be represented directly in a quantum state without complex preprocessing. The inputs to the quantum circuits are flowsheet representations expressed as binary connectivity matrices. Each element of the matrix indicates the presence or absence of a directed connection between unit operations. Because these variables are binary, they can be mapped to qubit states through basis encoding, where logical 0 and 1 correspond to the computational basis states $|0\rangle$ and $|1\rangle$, respectively. This avoids the need for any additional encoding steps.

The observables of the quantum circuit are used to approximate Q-function values associated with candidate actions. Since the Q-values do not correspond to physically interpretable quantities and are instead learned value estimates, there is no domain-specific decoding required. Measured expectation values are combined to form observables that are linearly scaled through trainable parameters to produce the final Q-value estimates for the discrete set of actions. As a result, the decoding step remains straightforward and does not introduce additional structural constraints which may be necessary for other applications. Under this formulation, RL-based process synthesis can be integrated with quantum RL, as both the state representation and value approximation can be expressed directly within a quantum circuit framework without complex feature engineering. 

\subsubsection{Algorithmic framework}
To assess the viability of quantum-based RL for process synthesis, the previously posed MDP problem must be solved with a quantum-enhanced algorithm. Towards this direction, the classical framework is reformulated by substituting the neural Q-function approximating components with the quantum counterparts (i.e., the NNs are replaced with quantum circuits), which is consistent with existing quantum-RL literature. The MDP definition, RL structure, and environment interaction loops are unchanged, which enables isolating the effect of Q-function approximation. 

Let $Q_{\theta}$ denote the parameterized state-action value function. In the classical benchmark, $\theta$ represents NN weights and biases. In the quantum-enhanced framework, $\theta$ instead represents the parameters of a quantum circuit, including rotation gate angles and scalable parameters. The underlying learning principles of RL remain unchanged since the objective of parameter updates is again to minimize the Bellman error with respect to the gradients defined by these tunable parameters. At each step of the algorithm, the compact matrix representation of the flowsheet is mapped to the inputs of the quantum circuit (e.g., qubit state initializations and rotation gate angle encodings). The circuit produces Q-estimates in a one-to-one mapping of each candidate action in the same action space as the classical approach. All hyperparameters, update scheduling, and reward function specifications are consistent in order to ensure a fair benchmarking comparison. 

The action selection and corresponding flowsheet modification, feasibility screening, and process simulation steps occur identically to the classical case. Screened flowsheets are simulated using the IDAES-PSE framework with continuous decision variables selected via IPOPT. In this way, the overall RL framework is structurally identical where only the Q-function approximation method is altered. The full Quantum-DQN algorithm designed to solve the process synthesis MDP is outlined in Algorithm 2. 

\begin{algorithm}[H]
\caption{Quantum-DQN for Process Synthesis}
    \begin{algorithmic}[1]
        \State Input $\mathcal{S}$ with size $n$, a user-defined set of candidate unit operations
        \State Initialize replay memory $\mathcal{B}$ with capacity $N$
        \State Initialize Q-PQC, $Q_{\theta}$, with random weights $\theta$
        \State Initialize target Q-PQC, $Q_{\theta^-}$ with weights $\theta^{-} = \theta$
        \For{episode $= 1$ \textbf{to} $M$}
            \State Initialize a blank flowsheet $s_0$
            \For{$t = 1$ \textbf{to} $T$}
                \State With probability $\epsilon$, select a random action $a_t$
                \State Otherwise select $a_t = \argmax_a Q_{\theta}(s_t, a)$
                \State Modify flowsheet according to action $a_t$ to generate flowsheet $s_{t+1}$
                \If{flowsheet $s_{t+1}$ passes screening}
                \State Simulate flowsheet to observe reward $r_t$
                \State Store transition $(s_t, a_t, r_t, s_{t+1})$ in $\mathcal{B}$
                \State Sample random batch of transitions $(s_j, a_j, r_j, s_{j+1})$ from $\mathcal{B}$
                \State Compute target action:
                \[
                a_{j+1} = \argmax_a Q_{\theta^-}(s_{j+1},a)
                \]
                \State Compute target:
                \[
                    y_j = r_j + \gamma \max_{a'} Q_{\theta^-}(s_{j+1}, a')
                \]
                \State Perform a gradient descent step on $(y_j - Q_{\theta}(s_j, a_j))^2$ with respect to $\theta$
                \Else{ Continue}
                \EndIf 
                \If{every $C$ steps}
                    \State Update target network: $\theta^{-} = \theta$
                \EndIf
                \State $s_t \gets s_{t+1}$
            \EndFor
        \EndFor
    \end{algorithmic}
\end{algorithm}

\subsubsection{State encoding strategies for algorithm improvements}
\subsubsection*{4.3.3.1 Variant 1}
The first quantum formulation employs a re-uploading PQC as the Q-function approximator. In this variant, the compact flowsheet representation is encoded directly into the qubit states, and the number of qubits is therefore equal to the dimension of the state representation. This algorithm was presented in our prior work \cite{braniffEnhancedReinforcementLearningdriven2025d}. The circuit follows a layered structure consisting of parameterized single-qubit rotations and entangling gates forming PQC layers. After each entangling block, the full classical state data is reintroduced through additional parameterized rotations as shown in Fig. \ref{fig:Reup_PQC}. The trainable parameters associated with both the rotation gates and the re-encoding operations are jointly optimized during training. Under this formulation, the Q-value approximation is entirely determined by the quantum state representation. No dimensionality reduction or classical preprocessing is applied prior to encoding. While this provides a direct mapping between the flowsheet matrix and the qubit initializations, it also implies that circuit width scales linearly with the size of the state space. As the number of candidate unit operations increases, the required qubit count and entanglement depth increase accordingly. This variant therefore serves as the most direct quantum analogue to the classical DQN formulation, differing only in the replacement of the neural value approximator with a fully state-encoded PQC. The algorithmic procedure remains identical to Algorithm 2, with the circuit architecture specified here.

\subsubsection*{4.3.3.2 Variant 2}
\begin{figure}[h]
    \centering
    \includegraphics[width=0.9\linewidth]{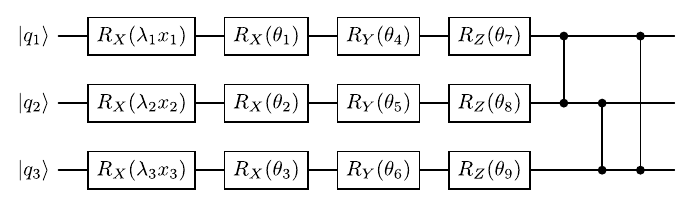}
    \caption{PQC using the encoding strategy described by Variant 2.}
    \label{fig:variant2}
\end{figure}

The second formulation modifies the state encoding strategy used in Variant 1. Rather than mapping the full flowsheet matrix directly onto individual qubits, the input features are instead injected into the circuit through parameterized rotation angles. This approach decouples the number of qubits required for the circuit from the dimensionality of the state representation, allowing the circuit width to be a design variable itself. In this variant, classical state features are introduced through Rx rotations for each qubit in each PQC layer following a sequential manner, where each feature is scaled by a trainable parameter prior to application as depicted in Fig. \ref{fig:variant2}. The encoded rotations occur first, followed by a PQC layer consisting of parameterized rotation and entangling gates. This sequence is repeated in multiple layers, allowing all features to be introduced in subsequent encoding steps. Because each qubit can encode one feature per layer through an Rx rotation, the total number of input features that can be represented depends on the circuit depth rather than the circuit width alone. An expression to find the minimum number of layers needed given the amount of qubits being used is provided below in Eq. \ref{eq:variant2_min}.

\begin{equation}
    \label{eq:variant2_min}
    L_{min} = \lceil \frac{n_x}{n_q} \rceil
\end{equation}
\noindent where $n_x$ is the number of input features to be encoded, $n_q$ is the number of qubits being used, $\lceil x \rceil$ represent the ceiling function, and $L_{min}$ is the minimum number of layers required to encode all inputs. As a result, the number of qubits can be selected independently of the state dimension, with additional encoding capacity obtained through layered data injection. The circuit depth can also be increased beyond the minimum number of layers required in order to increase expressibility or parameter counts. Under this structure, the Q-function approximation remains fully parameterized, but the representation of the state is distributed across circuit depth instead of being embedded directly into qubit count. All remaining elements of the reinforcement learning procedure follow Algorithm 2.

\subsubsection*{4.3.3.3 Variant 3}
\begin{figure}[h]
    \centering
    \includegraphics[width=0.9\linewidth]{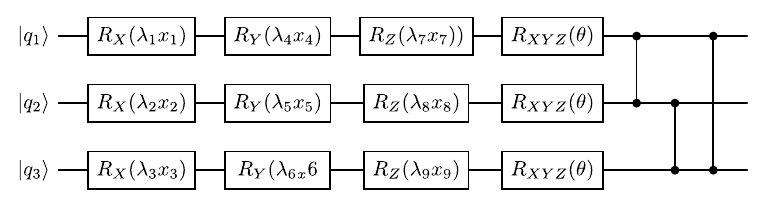}
    \caption{PQC using the encoding strategy described by Variant 3.}
    \label{fig:variant3}
\end{figure}

The third formulation extends the angle-based encoding strategy introduced in Variant 2. Rather than injecting input features exclusively through Rx rotations, this variant distributes encoded features across $R_x$, $R_y$ and $R_z$ rotations prior to the gate-based portion of each PQC layer. Within each encoding layer, classical features are mapped to rotation angles about the x, y, and z axes, with associated trainable scaling parameters, as shown in Fig. \ref{fig:variant3}. The encoded rotations are then followed by the same PQC operations used in the previous variants. As in Variant 2, there exists a minimum number of layers necessary to ensure all features can be incorporated when the number of input dimensions exceeds the per-layer encoding capacity. By utilizing all three rotational degrees of freedom, more features can be embedded per qubit within a single layer relative to the X-only formulation. Consequently, the required circuit depth for a given state dimension may be reduced while maintaining a fixed number of qubits. The total encoding capacity remains governed by the number of qubits and layers selected as shown below in Eq. \ref{eq:variant3_min}.

\begin{equation}
    \label{eq:variant3_min}
    L_{min} = \lceil \frac{n_x}{3n_q} \rceil
\end{equation}
\noindent where $n_x$ is the number of input features to be encoded, $n_q$ is the number of qubits being used, $\lceil x \rceil$ represents the ceiling function, and $L_{min}$ is the minimum number of layers required to encode all inputs. The total number of layers may be increased beyond what is necessary to increase parameter count and expressibility, however this comes with added computational cost when considering the quantum circuit. Compared to Variant 2, this encoding strategy is more efficient but comes with the cost of fewer tunable parameters when the minimum number of layers is used. 

To summarize, Variants 1–3 define a structured comparison of quantum state encoding strategies within the same reinforcement learning framework. Variant 1 encodes the full state directly into the quantum register, resulting in a circuit width that scales with the problem dimension. Variant 2 shifts this dependence from circuit width to circuit depth by injecting features through repeated single-axis rotations. Variant 3 further increases per-layer encoding capacity by utilizing rotations about multiple axes. In all cases, the reinforcement learning structure, reward formulation, and environment interaction remain unchanged. This controlled modification of the encoding mechanism enables a direct evaluation of how circuit width, depth, and feature injection strategy influence learning performance in process synthesis tasks.
\section{Case Study and Results Discussion}
\subsection{Problem description and setup}
This work considers a case study for the design of a hydrodealkylation (HDA) process adopted from \cite{wangReinforcementLearningAutomated2022}, in which toluene (C\textsubscript{6}H\textsubscript{5}CH\textsubscript{3}) reacts with hydrogen (H\textsubscript{2}) to produce benzene (C\textsubscript{6}H\textsubscript{6}) and methane (CH\textsubscript{4}) according to the reaction in Eq. \ref{eq:reaction}. 
\begin{equation}
\label{eq:reaction}
    C_6H_5CH_3 \ + \ H_2 \to C_6H_6 \ + \ CH_4
\end{equation}
The reaction is assumed to occur exclusively in the vapor phase without side reactions, providing a simplified setting for algorithmic evaluation. Feed conditions, summarized in Table \ref{tab:feedstreams}, consist of: (i) a vapor stream containing hydrogen and methane and (ii) a liquid stream containing toluene.

\begin{table}[!h]
\centering
\caption{Compositions and inlet conditions for feed streams.}
\label{tab:feedstreams}
\begin{tabular}{ccc}
\hline
 & Vapor Feed & Liquid Feed \\ 
\hline
Temperature (K) & 303.2 & 303.2 \\ 
\hline
Pressure (kPa) & 350 & 350 \\ 
\hline
\multicolumn{3}{c}{Flowrate (mol/s)} \\ 
\hline
Hydrogen & 0.30 & 0.00 \\ 
\hline
Toluene & 0.00 & 0.30 \\ 
\hline
Methane & 0.02 & 0.00 \\ 
\hline
Benzene & 0.00 & 0.00 \\ 
\hline
\end{tabular}
\end{table}

A maximum set of candidate unit operations is first selected that are available for flowsheet synthesis (such as heaters, coolers, stoichiometric reactors, etc.). The RL agent can then use any type and any number of the available unit operation(s) to build the flowsheet with any stream connections, thus resulting in a large combinatorial design space. Each unit operation is subject to operational constraints, such as maintaining outlet temperatures within specified ranges (e.g., 500–600 K). All phase separations are assumed to follow ideal vapor-liquid equilibrium behavior. The design objective is to identify flowsheet configuration(s) that maximize the vapor-phase benzene product, subject to the process specification that the benzene mole fraction in the final product must exceed 0.55. The reward function used in this experimental setup considers penalties for infeasibilities while rewarding higher purity and flowrate in the benzene product stream. For a more detailed analysis of the reward function, the reader is referred to the discussions in Tian et al. \cite{tianReinforcementLearningDrivenProcess2024d} and Wang et al. \cite{wangReinforcementLearningAutomated2022}.

Simulations are conducted using the IDAES-PSE environment, with continuous design and operating variables optimized by IPOPT for each proposed feasible flowsheet. This hybrid approach allows the RL agent to focus on discrete flowsheet decisions while relying on gradient-based optimization for continuous variables. To ensure reproducibility and a fair comparison across classical and quantum algorithms, all experiments share a common set of hyperparameters and training settings. All hyperparameters are summarized in Table A.1. By maintaining consistent training parameters and reward structures across algorithms, the comparison between classical and quantum RL approaches directly reflects differences in value-function representation and encoding strategies, rather than differences in training protocol. This controlled setup forms the core benchmark of the study.

\begin{figure}[!h]
    \centering
    \includegraphics[width=0.8\linewidth]{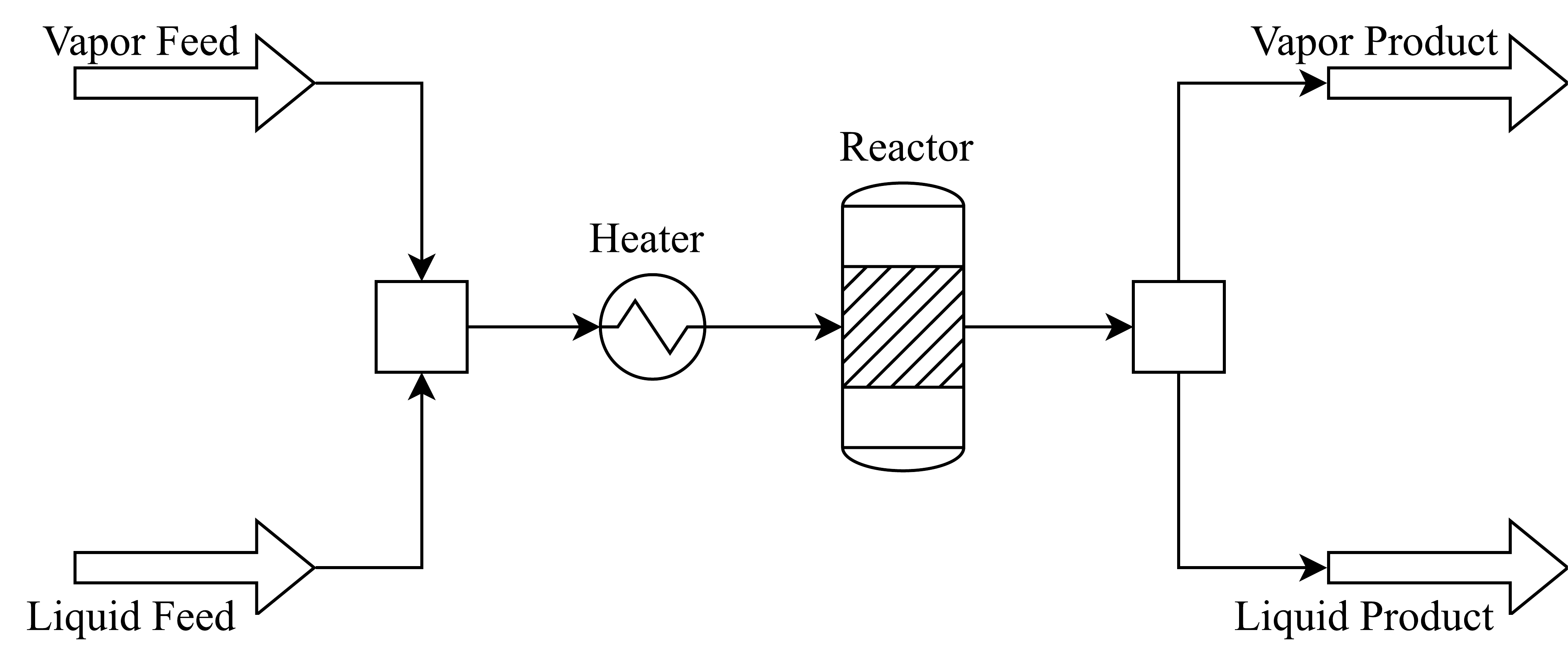}
    \caption{Optimal flowsheet configuration with maximum reward.}
    \label{fig:flowsheet}
\end{figure}

Three design scenarios are considered. Scenario 1 evaluates a small restricted problem considering only 2 candidate unit operations. Scenarios 2 and 3 examine increasingly larger design spaces, providing an opportunity to explore how circuit architecture and encoding strategies influence algorithmic performance. The goal of each scenario is to generate the optimal flowsheet design for this process subject to the operational constraints. Despite the many possible flowsheet designs, the optimal process design across all scenarios only requires a single reactor and a heater, as seen in Fig. \ref{fig:flowsheet}. This design has a benzene flowrate of 0.225 mol/s with a purity of 0.75, giving a maximum instantaneous reward of 1350. 

\subsection{Scenario 1}
This scenario examines the simplest HDA process design, considering only two candidate unit operations, a reactor and a heater. The agent can select any one, or both, of these two units with any stream connections. This configuration represents the minimal subset of units required to achieve the optimal flowsheet for the target benzene product. All reinforcement learning algorithms, the classical DQN and the three quantum variants, were trained for 30,000 episodes. Hyperparameters are consistent with those listed in Table A.1, including reward structure, optimizer settings, and initial exploration rates. Each algorithm successfully identifies the optimal flowsheet (i.e., Fig. \ref{fig:flowsheet}), demonstrating that even the simplest quantum-enhanced approaches can match the performance of classical DQN in this minimal problem setting. Performance metrics, including the number of times the optimal solution was found (Opt. SF), the number of unique solutions found (Uniq. SF), the number of feasible solutions found (Feas. SF), the first episode where the optimal solution was found ($1^{st}$ Opt. Ep.) and the total runtime in hours, are summarized in Table \ref{tab:scenario1} for a total of 10 random runs for each algorithm while the comparative modeling information for each algorithm is given in Table \ref{tab:scenario1_models}. The information for seed reproducibility is reported in Appendix B, and the details of classical DQN architecture are presented in Appendix C. The itemized performance for each run is reported in Table D.1, Appendix D.

\begin{table}[!h]
\centering
\caption{Performance over 10 random runs (mean $\pm$ std) for Scenario 1.}
\label{tab:scenario1}
\begin{tabular}{cccccc}
\toprule
Method &Opt. SF &Uniq. SF &Feas. SF &
$1^{st}$ Opt. Ep. &Runtime (hr) \\
\midrule
Classical &169 $\pm$ 21 &3 $\pm$ 0 &1 $\pm$ 0 &
244 $\pm$ 267 &0.062 $\pm$ 0.002 \\

Variant 1 &113 $\pm$ 15 &3 $\pm$ 0 &1 $\pm$ 0 &
219 $\pm$ 153 &1.190 $\pm$ 0.241 \\

Variant 2 &110 $\pm$ 13 &3 $\pm$ 0 &
1 $\pm$ 0 &120 $\pm$ 95 &1.168 $\pm$ 0.100 \\

Variant 3 &113 $\pm$ 10 &3 $\pm$ 0 &1 $\pm$ 0 &
273 $\pm$ 235 &0.628 $\pm$ 0.070 \\
\bottomrule
\end{tabular}
\end{table}

\begin{table}[!h]
\centering
\caption{Model architecture information for Scenario 1.}
\label{tab:scenario1_models}
\begin{tabular}{lccc}
\toprule
Method &Trainable Parameters &Qubits &Layers \\
\midrule
Classical & 2180 & -- & 1 \\
Variant 1 & 88 & 12 & 1 \\
Variant 2 & 64 & 4 & 3 \\
Variant 3 & 40 & 4 & 1 \\
\bottomrule
\end{tabular}
\end{table}

\subsection{Scenario 2}
This scenario considers three candidate unit operations: a reactor, a heater, and a heat exchanger. The agent can select any one, two, or three of these units with any stream connections. This setup introduces additional combinatorial complexity compared to Scenario 1, while remaining small enough for efficient evaluation across all reinforcement learning algorithms. All algorithms, except for Variant 1 of the quantum approach, were trained for 30,000 episodes using the hyperparameters defined in Table A.1. At this level of complexity, 20 qubits are required for the direct one-to-one mapping used in Variant 1, whereas only 4 qubits are used for the other variants. With this many qubits needed for Variant 1, it becomes infeasible to simulate in a tractable amount of time, taking an estimated 106 hours when extrapolating from running this variant for 10 episodes. The performance metrics for each algorithm applied to Scenario 2 are given below in Table \ref{tab:scenario2} with modeling information provided in Table \ref{tab:scenario2_models}. Each set of 10 runs for each algorithm utilizes the same set of pseudorandom numbers for fair algorithmic comparison. The information for seed reproducibility is reported in Appendix B, and the itemized performance for each run is reported in Table D.2, Appendix D. The reward function, state representation, and environment interaction remain consistent with the setup described in Section 5.1, ensuring a controlled benchmark comparison. Across all approaches, the optimal flowsheet is reliably identified, demonstrating that the quantum variants maintain performance with classical DQN while handling slightly increased combinatorial complexity.

\begin{table}[!h]
\centering
\caption{Performance over 10 random runs (mean $\pm$ std) for Scenario 2.}
\label{tab:scenario2}
\begin{tabular}{cccccc}
\toprule
Method &Opt. SF &Uniq. SF &Feas. SF &
$1^{st}$ Opt. Ep. &Runtime (hr) \\
\midrule
Classical &14 $\pm$ 5 &11 $\pm$ 0 &1 $\pm$ 0 &2152 $\pm$ 1766 &0.059 $\pm$ 0.005 \\
Variant 1 &-- &-- &-- &-- &$\approx$106 \\

Variant 2 &8 $\pm$ 3 &11 $\pm$ 0 &1 $\pm$ 0 &
3044 $\pm$ 3243 &1.463 $\pm$ 0.278 \\

Variant 3 &8 $\pm$ 2 &11 $\pm$ 0 &
1 $\pm$ 0 &2361 $\pm$ 2942 &1.067 $\pm$ 0.186 \\
\bottomrule
\end{tabular}
\end{table}

\begin{table}[!h]
\centering
\caption{Model architecture information for Scenario 2.}
\label{tab:scenario2_models}
\begin{tabular}{lccc}
\toprule
Method &
Trainable Parameters &
Qubits &
Layers \\
\midrule
Classical & 3333 & -- & 1 \\
Variant 1 & 145 & 20 & 1 \\
Variant 2 & 100 & 5 & 4 \\
Variant 3 & 80 & 5 & 2 \\
\bottomrule
\end{tabular}
\end{table}

\subsection{Scenario 3}
This scenario evaluates a larger HDA process design problem with four candidate unit operations: a reactor, a heater, a cooler, and a heat exchanger. The increased number of units significantly expands the combinatorial design space, providing an opportunity to examine the performance of classical and quantum reinforcement learning algorithms under more challenging conditions. All algorithms except for Variant 1 were trained for 60,000 episodes to ensure convergence and allow sufficient exploration of the design space. The episode count is increased because each algorithm was unable to find the max-reward solution within the first 30,000 episodes consistently. For this problem, Variant 1 needs 30 qubits in its quantum circuit which uses more memory than is available for use in the classical computer. For this reason, Variant 1 is fully excluded from this scenario. Additionally, this scenario introduces parameter-boosted (PB) runs for Variants 2 and 3 to investigate if the performance of the quantum algorithms is constrained by model expressibility. The results, summarized in Tables \ref{tab:scenario3}--\ref{tab:scenario3_models}, illustrate the key differences between approaches. While all algorithms are capable of identifying feasible flowsheets, quantum Variants 2 and 3 demonstrate improved efficiency in exploring the design space when controlling for parameter counts. These observations highlight the impact of encoding strategies and circuit architecture when addressing larger, more complex process synthesis problems.

\begin{table}[!h]
\centering
\caption{Performance over a single random run for Scenario 3.}
\label{tab:scenario3}
\begin{tabular}{cccccc}
\toprule
Method &Opt. SF &Uniq. SF &Feas. SF &
$1^{st}$ Opt. Ep. &Runtime (hr) \\
\midrule
Classical &6 &31 &1&11255 &0.085 \\
Variant 1 &-- &-- &-- &-- &--\\
Variant 2 &1 &19 &2&30815 &3.040 \\
Variant 3 &0 &19 &0 &-- &2.315 \\
Variant 2 (PB) &11 &29 &2&4105 &13.146 \\
Variant 3 (PB) &9 &30 &2 &8296 &11.539 \\
\bottomrule
\end{tabular}
\end{table}

\begin{table}[!h]
\centering
\caption{Model architecture information for Scenario 3.}
\label{tab:scenario3_models}
\begin{tabular}{lccc}
\toprule
Method &
Trainable Parameters &
Qubits &
Layers \\
\midrule
Classical & 4742 & -- & 1 \\
Variant 1 & 216 & 30 & 1 \\
Variant 2 & 100 & 6 & 5 \\
Variant 3 & 80 & 6 & 2 \\
Variant 2 (PB) & 744 & 6 & 30 \\
Variant 3 (PB) & 744 & 6 & 20 \\
\bottomrule
\end{tabular}
\end{table}

\subsection{Comparative Performance Analysis}
The three scenarios collectively provide a structured benchmark for evaluating classical and quantum reinforcement learning approaches under progressively increasing combinatorial complexity. Because the underlying MDP, reward formulation, and training protocol remain fixed across experiments, differences in performance can be attributed primarily to the value-function representation and encoding strategy. Across all scenarios, the quantum variants utilize substantially fewer tunable parameters than the classical DQN benchmark. While the computational time for the classical approach is significantly less than all quantum algorithms, it is important to note that the quantum algorithms were implemented on a classical computer simulating a quantum machine. The computational times reported are meant to be used as a comparison between the quantum algorithms and do demonstrate their scalability in complexity. It is impossible to determine the time these algorithms would take without implementing them on real quantum hardware. 

In Scenario 1, all approaches reliably identify the optimal flowsheet multiple times during training (Table. \ref{tab:scenario1}), indicating that for small design spaces the representational capacity of both neural and parameterized quantum models is sufficient. In Scenario 2, differences between the algorithms begin to emerge. While all algorithms reach the maximum reward solution, the quantum variants have a decreased performance in exploration behavior and convergence frequency. This is likely explained by the difference in tunable parameters. While the quantum variants use an order of magnitude fewer parameters (Table \ref{tab:scenario2_models}), they are able to sufficiently identify the optimal flowsheet within the episode limit. However, on average, the quantum variants do so slower than the classical algorithm (Table \ref{tab:scenario2}).

Scenario 3 reveals clearer distinctions between all approaches. The classical DQN uses 4,742 tunable parameters (Table \ref{tab:scenario3_models}) and demonstrates stronger consistency in rediscovering the optimal flowsheet over the full training horizon versus the minimal-layer implementations of Variants 2 and 3. However, when increasing the number of tunable parameters for quantum Variants 2 and 3, a different trend emerges. Using 744 tunable parameters each, the PB quantum formulations achieve a slightly higher frequency of optimal solutions and are able to find an additional feasible flowsheet (Table \ref{tab:scenario3}). The PB quantum agents still maintain almost seven times fewer parameters than the classical DQN but perform slightly better on per-episode and per-parameter bases. This is further supported by the fact that the PB quantum agents discover the optimal solution quicker than the classical approach (Table \ref{tab:scenario3}), however this claim is limited given that Scenario 3 examines a single run only.

A comparison among the quantum variants further clarifies the role of encoding. The encoding strategy in Variant 2 typically outperforms the multi-axis encoding strategy in Variant 3 throughout all scenarios. This is potentially due to the fact that Variant 2 requires additional circuit depth to encode the same number of features. This result suggests that increased parameter and layer count may contribute more to performance than per-layer encoding density. This is supported by the results in Table \ref{tab:scenario3} which show when Variants 2 and 3 have enough layers to provide a comparable amount of parameters, they have similar but not identical performances. 

Overall, the results demonstrate that:
\begin{itemize}
    \item For small design spaces, classical and quantum approaches achieve similar outcomes.
    \item As combinatorial complexity increases, architectural choices in the value-function approximator become increasingly influential, particularly when the qubit amount is a limiting factor. 
    \item Parameter efficiency alone does not fully explain performance differences. Circuit structure and encoding strategy also play a role.
    \item Despite the potential of the proposed approach, we acknowledge that the problems investigated in this work are still at small scale compared to typical flowsheet design problems. Continued investigation into quantum circuit design for reinforcement learning is a necessity, particularly for combinatorial design problems.
\end{itemize}

\section{Conclusions}

This work presented a structured benchmark study comparing classical and quantum value-function approximators within a reinforcement learning framework for process synthesis. A consistent MDP formulation and training protocol were used across all experiments, enabling a controlled evaluation of how representation choice influences learning behavior in combinatorial design problems. Three quantum circuit formulations were investigated and compared against a classical deep Q-network benchmark across progressively larger HDA process design scenarios. For small design spaces, all approaches reliably identified the optimal flowsheet, indicating that both neural and variational quantum models possess sufficient representational capacity under limited combinatorial complexity. In larger design spaces the classical DQN consistently exhibits strong performance, but when parameter counts across models are accounted for, the quantum-enhanced approach performs slightly better. Among the quantum variants, encoding strategy and circuit depth were observed to influence exploration ability and solution quality, particularly in the most complex scenario. Overall, this study presents a systematic MDP formulation for process synthesis problems and establishes a reproducible framework for benchmarking quantum reinforcement learning methods in process synthesis. The results are consistent with the hypothesis that quantum value-function approximators can match classical DQN performance at lower parameter counts, but rigorous comparisons await hardware deployment. The findings provide insight into how circuit architecture, encoding strategy, and model capacity interact when addressing discrete decision-making problems. The current study is limited up to a 4-unit design problem, and while this is useful for controlled benchmarking, it is not yet complex enough to deliver broad claims for all scales of combinatorial problems. Future work will focus on expanding benchmark problems to larger design spaces, investigating alternative circuit architectures, and improving training efficiency. Continued development of hardware-aware quantum reinforcement learning strategies will be necessary to assess practical deployment on emerging quantum platforms. 

\section*{Acknowledgments}
The authors acknowledge financial support from NSF EPSCoR RII Track-4 OIA-2327303, NSF GRFP 2024370240, and Department of Chemical and Biomedical Engineering at West Virginia University.

\appendix
\newpage
\section{Hyperparameters}
The hyperparameters used in this work were selected based on the classical benchmark algorithm adopted from Wang et al. \cite{wangComprehensiveReviewQuantum2024}. These hyperparameters were held constant across all algorithms and scenarios investigated and are summarized in Table A.1. 
\begin{table}[!h]
\centering
\caption*{Table A.1: Values for the hyperparameters used within each algorithm.}
\label{tab:hyperparameters}
\begin{tabular}{|c|c|}
\hline
Hyperparameter & Value \\ 
\hline
Learning Rate & 0.01 \\ 
\hline
Discount Factor ($\gamma$) & 0.5  \\ 
\hline
Batch Size & 32  \\ 
\hline
Replay Buffer Size & 20000 \\ 
\hline
Target Network Update Frequency & 200 \\ 
\hline
Exploration Rate ($\epsilon$) & 0.08 \\ 
\hline
Exploration Decay Rate &  0.01\\ 
\hline
\end{tabular}
\end{table}
\section{Computing Environment}
All experiments were conducted on an Alienware m16
laptop with 32 GB RAM, an Intel i9-13900HX CPU, and an NVIDIA GeForce RTX 4090 Laptop GPU. All computational experiments were conducted using Python 3.9 within the Windows Subsystem for Linux (WSL) environment. Quantum circuit simulations were implemented using Cirq v1.3.0, while machine learning models were developed and trained using TensorFlow Quantum v0.7.2 and TensorFlow v2.15.0. For reproducibility, the first run of the ten runs for Scenarios 1 and 2 was initialized using a fixed random seed (seed = 1) for all algorithms. Subsequent runs within the batch followed the resulting sequence generated during training and exploration, which contained an identical amount of pseudorandom number calls, allowing all runs to be fairly compared.

\section{Classical DQN Architecture}
The classical DQN consisted of a single dense hidden layer with 128 nodes using the ReLU activation function with input and output layers meeting problem-specific dimensions. The specific architecture dimensions are outlined in Table C.1 for each scenario reported in Section 5.

\begin{table}[!h]
\centering
\caption{Layer dimensions of the classical DQN for each scenario.}
\label{tab:DQN_architecture}
\begin{tabular}{lccc}
\toprule
Layer & Scenario 1 & Scenario 2 & Scenario 3 \\
\midrule
Input & 12 & 20 & 30 \\
Hidden & 128 & 128 & 128 \\
Output & 4 & 5 & 6 \\
\bottomrule
\end{tabular}
\end{table}

\begin{landscape}
\section{Performance for Scenarios 1 and 2}
This appendix presents the performance for each random run in Scenarios 1 and 2. 

\begin{table}[htbp!]
    \centering
    \caption*{Table D.1: Performance for each random run in Scenario 1.}
    \adjustbox{max width = 1.4\textwidth}{
    \begin{tabular}{l | l | rrrrrrrrrr | rr}
    \hline
     & Run & 1 & 2 & 3 & 4 & 5 & 6 & 7 & 8 & 9 & 10 & Mean & SD\\
    \hline
    \multirow{5}{*}{Classical} & Number of max reward solutions	&	172	&	195	&	187	&	163	&	127	&	164	&	197	&	152	&	172	&	156	&	169	&	21	\\
    & Unique Solutions	&	3	&	3	&	3	&	3	&	3	&	3	&	3	&	3	&	3	&	3	&	3	&	0	\\
    & Feasible Solutions	&	1	&	1	&	1	&	1	&	1	&	1	&	1	&	1	&	1	&	1	&	1	&	0	\\
    & Episode where optimal first found	&	64	&	51	&	37	&	946	&	217	&	61	&	276	&	287	&	246	&	258	&	244	&	267	\\
    & Total run time (hr)	&	0.058	&	0.058	&	0.062	&	0.064	&	0.063	&	0.064	&	0.063	&	0.063	&	0.065	&	0.063	&	0.062	&	0.002	\\
    \hline
    \multirow{5}{*}{Variant 1} & Number of max reward solutions	&	146	&	123	&	105	&	114	&	98	&	114	&	99	&	123	&	104	&	99	&	113	&	15	\\
    & Unique Solutions	&	3	&	3	&	3	&	3	&	3	&	3	&	3	&	3	&	3	&	3	&	3	&	0	\\ 
    & Feasible Solutions	&	1	&	1	&	1	&	1	&	1	&	1	&	1	&	1	&	1	&	1	&	1	&	0	\\ 
    & Episode where optimal first found	&	64	&	160	&	329	&	336	&	103	&	134	&	519	&	122	&	80	&	349	&	219	&	153	\\
    & Total run time (hr)	&	0.761	&	0.890	&	1.033	&	1.087	&	1.195	&	1.437	&	1.465	&	1.312	&	1.290	&	1.425	&	1.190	&	0.241	\\
    \hline
    \multirow{5}{*}{Variant 2} & Number of max reward solutions	&	125	&	125	&	96	&	105	&	94	&	94	&	125	&	114	&	119	&	103	&	110	&	13	\\
    & Unique Solutions	&	3	&	3	&	3	&	3	&	3	&	3	&	3	&	3	&	3	&	3	&	3	&	0	\\
    & Feasible Solutions	&	1	&	1	&	1	&	1	&	1	&	1	&	1	&	1	&	1	&	1	&	1	&	0	\\
    & Episode where optimal first found	&	64	&	10	&	49	&	270	&	186	&	191	&	243	&	72	&	15	&	96	&	120	&	95	\\
    & Total run time (hr)	&	0.960	&	1.118	&	1.075	&	1.174	&	1.179	&	1.138	&	1.219	&	1.245	&	1.306	&	1.262	&	1.168	&	0.100	\\
    \hline
    \multirow{5}{*}{Variant 3} & Number of max reward sols	&	124	&	125	&	109	&	108	&	103	&	99	&	115	&	128	&	122	&	104	&	113	&	10	\\
    & Unique Solutions	&	3	&	3	&	3	&	3	&	3	&	3	&	3	&	3	&	3	&	3	&	3	&	0	\\
    & Feasible Solutions	&	1	&	1	&	1	&	1	&	1	&	1	&	1	&	1	&	1	&	1	&	1	&	0	\\
    & Episode where optimal first found	&	64	&	39	&	160	&	217	&	571	&	140	&	168	&	126	&	622	&	621	&	273	&	235	\\
    & Total run time (hr)	&	0.473	&	0.547	&	0.611	&	0.623	&	0.700	&	0.669	&	0.657	&	0.655	&	0.658	&	0.687	&	0.628	&	0.070	\\
    \hline
    \multicolumn{14}{l}{* SD: Standard Deviation}
    \end{tabular}}
    \label{tab:D1}
\end{table}
\end{landscape}

\begin{landscape}
\begin{table}[htbp!]
    \centering
    \caption*{Table D.2: Performance for each random run in Scenario 2.}
    \adjustbox{max width = 1.4\textwidth}{
    \begin{tabular}{l | l | rrrrrrrrrr | rr}
    \hline
     & Run & 1 & 2 & 3 & 4 & 5 & 6 & 7 & 8 & 9 & 10 & Mean & SD\\
    \hline
    \multirow{5}{*}{Classical} & Number of max reward solutions	&	16	&	15	&	20	&	13	&	10	&	13	&	12	&	22	&	5	&	16	&	14	&	5	\\
    & Unique Solutions	&	11	&	11	&	11	&	11	&	11	&	11	&	11	&	11	&	11	&	11	&	11	&	0	\\
    & Feasible Solutions	&	1	&	1	&	1	&	1	&	1	&	1	&	1	&	1	&	1	&	1	&	1	&	0	\\
    & Episode where optimal first found	&	1655	&	28	&	73	&	1528	&	2577	&	387	&	4222	&	2185	&	3884	&	4981	&	2152	&	1766	\\
    & Total run time (hr)	&	0.063	&	0.063	&	0.063	&	0.045	&	0.061	&	0.059	&	0.060	&	0.060	&	0.059	&	0.056	&	0.059	&	0.005	\\
    \hline
    \multirow{5}{*}{Variant 2} & Number of max reward solutions	&	13	&	9	&	6	&	10	&	5	&	5	&	7	&	12	&	5	&	7	&	8	&	3	\\
    & Unique Solutions	&	11	&	11	&	11	&	11	&	11	&	11	&	11	&	11	&	11	&	11	&	11	&	0	\\
    & Feasible Solutions	&	1	&	1	&	1	&	1	&	1	&	1	&	1	&	1	&	1	&	1	&	1	&	0	\\
    & Episode where optimal first found	&	1151	&	1366	&	10767	&	1926	&	2274	&	2603	&	584	&	2005	&	6888	&	876	&	3044	&	3243	\\
    & Total run time (hr)	&	1.010	&	1.288	&	1.498	&	1.497	&	1.794	&	1.559	&	1.949	&	1.190	&	1.309	&	1.533	&	1.463	&	0.278	\\
    \hline
    \multirow{5}{*}{Variant 3} & Number of max reward solutions	&	12	&	9	&	6	&	9	&	6	&	7	&	8	&	7	&	8	&	6	&	8	&	2	\\
    & Unique Solutions	&	11	&	11	&	11	&	11	&	11	&	11	&	11	&	11	&	11	&	11	&	11	&	0	\\
    & Feasible Solutions	&	1	&	1	&	1	&	1	&	1	&	1	&	1	&	1	&	1	&	1	&	1	&	0	\\
    & Episode where optimal first found	&	1326	&	853	&	1749	&	351	&	302	&	2573	&	1526	&	2234	&	2266	&	10433	&	2361	&	2942	\\
    & Total run time (hr)	&	0.713	&	0.919	&	0.878	&	1.059	&	1.086	&	1.079	&	1.143	&	1.220	&	1.279	&	1.298	&	1.067	&	0.186	\\
    \hline
    \end{tabular}}
    \label{tab:D2}
\end{table}
\end{landscape}





\newpage
\bibliographystyle{elsarticle-num} 
\bibliography{bibliography}
\end{document}